\documentclass[10pt]{article}
\usepackage{graphicx}  
\usepackage[english]{babel}
\usepackage{bm}        
\usepackage{amssymb}   
\usepackage{amsmath}   
\usepackage{url}
\usepackage{siunitx}
\usepackage{braket}
\usepackage{times}
\usepackage{lineno}
\usepackage{scicite}
\usepackage{color}     
\usepackage[]{todonotes} 
\usepackage{soul}

\newenvironment{sciabstract}{%
\begin{quote} \bf}
{\end{quote}}

\title{
Quantum Gas Mixtures and Dual-Species Atom Interferometry in Space}

\author {Ethan R. Elliott,$^{1\ast}$ David C. Aveline,$^{1}$ Nicholas P. Bigelow,$^{2}$\\Patrick Boegel,$^{3}$ Sofia Botsi,$^{1}$ Eric Charron,$^{4}$ Jos\'e P. D'Incao, $^{5}$
Peter Engels,$^{6}$ \\Timothé Estrampes,$^{4,7}$ Naceur Gaaloul,$^{7}$ James R. Kellogg,$^{1}$ James M. Kohel,$^{1}$ \\Norman E. Lay,$^{1}$ Nathan Lundblad,$^{8}$ Matthias Meister,$^{9}$ Maren E. Mossman,$^{6,10}$ \\Gabriel M\"{u}ller,$^{7}$ Holger M\"{u}ller,$^{11}$ Kamal Oudrhiri,$^{1}$ Leah E. Phillips,$^{1}$ \\Annie Pichery,$^{4,7}$ Ernst M. Rasel,$^{7}$ Charles A. Sackett,$^{12}$ Matteo Sbroscia,$^{1}$\\ Wolfgang P. Schleich,$^{3,13,14,15}$ Robert J. Thompson,$^{1}$ Jason R. Williams$^{1\ast}$ \\
\normalsize{$^{1}$Jet Propulsion Laboratory, California Institute of Technology, Pasadena, CA 91109, USA}\\
\normalsize{$^{2}$Department of Physics and Astronomy, University of Rochester, Rochester, NY, 14627, USA}\\
\normalsize{$^{3}$Institut f\"{u}r Quantenphysik and Center for Integrated Quantum Science and Technology (IQST),}\\
\normalsize{Ulm University, Ulm, Germany}\\
\normalsize{$^{4}$Université Paris-Saclay, CNRS, Institut des Sciences Moléculaires d’Orsay, F-91405 Orsay, France}\\
\normalsize{$^{5}$JILA, NIST, and the Department of Physics, University of Colorado, Boulder, CO 80309, USA}\\
\normalsize{$^{6}$Department of Physics and Astronomy, Washington State University, Pullman, WA, USA 99164}\\
\normalsize{$^{7}$Leibniz University Hannover, Institute of Quantum Optics, }\\ 
\normalsize{QUEST-Leibniz Research School, Hanover, Germany}\\
\normalsize{$^{8}$Department of Physics and Astronomy, Bates College, Lewiston, ME, USA}\\
\normalsize{$^{9}$German Aerospace Center (DLR), Institute of Quantum Technologies, Ulm, Germany}\\
\normalsize{$^{10}$Department of Physics and Biophysics, University of San Diego, San Diego, CA, USA 92110}\\
\normalsize{$^{11}$Department of Physics, University of California, Berkeley, CA, USA}\\
\normalsize{$^{12}$Physics Department, University of Virginia, Charlottesville, Virginia 22904, USA}\\
\normalsize{$^{13}$Hagler Institute for Advanced Study, Texas A\&M University, College Station, TX, USA}\\
\normalsize{$^{14}$Texas A\&M AgriLife Research, Texas A\&M University, College Station, TX, USA}\\
\normalsize{$^{15}$Institute for Quantum Science and Engineering (IQSE), Department of Physics and Astronomy,}\\
\normalsize{Texas A\&M University, College Station, TX, USA}\\
\normalsize{$^\ast$Corresponding Authors: Ethan.R.Elliott@jpl.nasa.gov, Jason.R.Williams.Dr@jpl.nasa.gov}\\
}

\date{\today}
\begin{document}
\maketitle
\begin{sciabstract}

The capability to reach ultracold atomic temperatures in compact instruments has recently been extended into space~\cite{Becker2018,CAL_nature_20}. Ultracold temperatures amplify quantum effects, while free-fall allows further cooling and longer interactions time with gravity - the final force without a quantum description. On Earth, these devices have produced macroscopic quantum phenomena such as Bose-Einstein condensation (BECs), superfluidity, and strongly interacting quantum gases~\cite{Levin2012UltracoldBA}. Quantum sensors interfering the superposition of two ultracold atomic isotopes have tested the Universality of Free Fall (UFF), a core tenet of Einstein’s classical gravitational theory, at the $10^{-12}$ level~\cite{AI_Kasevich_2022}. In space, cooling the elements needed to explore the rich physics of strong interactions and preparing the multiple species required for quantum tests of the UFF has remained elusive. Here, utilizing upgraded capabilities of the multi-user Cold Atom Lab (CAL) instrument within the International Space Station (ISS), we report the first simultaneous production of a dual species Bose-Einstein condensate in space (formed from $^{87}$Rb and $^{41}$K), observation of interspecies interactions, as well as the production of $^{39}$K ultracold gases. We have further achieved the first space-borne demonstration of simultaneous atom interferometry with two atomic species ($^{87}$Rb and $^{41}$K). These results are an important step towards quantum tests of UFF in space, and will allow scientists to investigate aspects of few-body physics, quantum chemistry, and fundamental physics in novel regimes without the perturbing asymmetry of gravity.
\end{sciabstract}

It is a period of rapid expansion and development for space-based platforms exploring quantum science and fundamental physics with ultracold atoms. A sub-orbital rocket payload has created Bose-Einstein condensates (BECs) and performed atom interferometry in space~\cite{Becker2018,space_AI}. 
Under remote operation from NASA’s Jet Propulsion Laboratory (JPL), the multi-user Cold Atom Lab (CAL) has performed daily experiments based on $^{87}$Rb BECs for the last five years onboard the International Space Station (ISS). Hardware support, provided by the NASA astronaut corps, has recently included two significant CAL upgrades to achieve matter-wave interferometry and sympathetic cooling of bosonic potassium. 
On Earth, future cold atom devices bound for space are under active development, integration, and testing, including the Bose-Einstein Condensate and Cold Atom Laboratory (BECCAL)~\cite{Frye2021}, the Cold Atom Physics Research Rack (CAPR)~\cite{10098699}, and the next MAIUS missions~\cite{elsen2023dualspecies}.
This growing infrastructure of space-based ultracold atomic physics has the potential for transformative discoveries in gravity science, cosmology, the nature of dark energy and dark matter, and quantum simulations of interacting many-body systems~\cite{Safronova18,Bassi2022}. Central to these prospects is the remarkable interplay between ultracold temperatures, quantum phenomena, and processes enhanced by the microgravity conditions of space. For each benefit of microgravity, there are new capabilities provided by the binary atomic mixtures described here.

Ultracold atomic systems, realized through laser-cooling and trapping techniques~\cite{Raa87,Phi82}, are broadly used throughout the physical sciences~\cite{Blo12, Bra06, Sal12} for two key reasons: increased control over atomic behavior and the amplification of quantum effects. The quantum wavelength of a massive particle is inversely related to its momentum, so by cooling an atom to temperatures within a millionth of a degree of absolute zero, the reduction in momentum causes the quantum wave to extend to micrometer length scales. This is 
strikingly seen in the creation of large quantum objects such as the Bose-Einstein condensate~\cite{And95,davis1995PRL} or the closely related phenomenon of superfluidity~\cite{greiner2002quantum}. Maturation of the unique hardware at the core of ultracold atomic experiments has enabled deployment in space, achieving reliable remote operation and surviving harsh launch conditions~\cite{Becker2018, CAL_nature_20}. 
In microgravity, the cooling techniques of adiabatic decompression~\cite{Lean03} and delta-kick collimation~\cite{DKC_New} can be enhanced to create new regimes of ultra-low temperatures, densities, and free-space expansion rates~\cite{Gaaloul2022,PhysRevLett.127.100401,PhysRevLett.114.143004}. Given the appropriate atomic species and mixtures, these conditions are ideal for studying the emergence of complexity from controllable atomic systems, including quantum bubbles \cite{PhysRevA.106.013309} and the formation of
large and delicate quantum objects, such as Feshbach molecules~\cite{chin2010rmp} or Efimov trimers~\cite{chapurin2019prl,PhysRevLett.125.243401}. Further, a free-falling experimental apparatus extends the time that an unconfined quantum object has to interact with gravity before colliding with a surface of the device. These benefits compound for ultracold atom interferometers (AI), which utilize light pulses to manipulate matter waves into an interfering quantum superposition where the resulting phase difference is affected by inertial forces like gravity. The sensitivity of AI can scale with the square of time between light pulses~\cite{Safronova18}, a duration increased by slow expansion rates and extended free fall. A binary-BEC system in space allows an interferometer to simultaneously probe the interaction of two quantum test masses with gravity, a critical step towards a quantum test of the theory of General Relativity in space. 

Einstein’s classical theory of General Relativity, our best description of gravity, is based on the Weak Equivalence Principle, or the Universality of Free Fall (UFF). The UFF requires that the trajectory of an object in space-time is independent of its composition. While the legacy of experimentally testing this idea reaches back to the Renaissance, modern interest is sustained by theoretical attempts to reconcile quantum mechanics and gravity that lead to small UFF violations. Classically, this can be tested by measuring the differential acceleration between two test-masses of different structure (i.e. a boulder and a pebble), while the AI analogue can use isotopically pure quantum test masses of different atomic species. On Earth, the most advanced AI test of the UFF compared isotopes of $^{87}$Rb and $^{85}$Rb using a 10 m high vacuum chamber to achieve 2~s of free fall, and found no violation at the level of $10^{-12}$~\cite{AI_Kasevich_2022}. Recently, a $10^{-15}$ level of accuracy was achieved by a classical experiment within a satellite, capitalizing on the advantages of operating in space for precision UFF tests~\cite{micro2022}. Owing to the fundamental differences in the description of classical and quantum motion, combining these two research paths by studying multiple quantum test masses in space is a long-standing goal that has led to a variety of proposed missions~\cite{STE_QUEST,stequest2022,QTEST,Frye2021,guage_2009}. 

CAL currently serves as a multi\-user facility to support the work of five science teams~\cite{Bigelow2015TB,Cornell2017TB,Williams2017TB,Lundblad2016TB,Sackett2016TB}. Prior to the results reported in this manuscript, space-borne experiments with BECs, adiabatic decompression, AI, and unique microgravity topologies have relied solely on the atomic species $^{87}$Rb~\cite{Becker2018,CAL_nature_20,Sackett2022,Carollo2022,Gaaloul2022,Williams2022}. As an atom, $^{87}$Rb features well-resolved hyperfine states which make it receptive to established cooling techniques that can be implemented with easily accessible, reliable, and inexpensive laser systems.
However, richer physics becomes accessible with the introduction of more diverse atomic species. 
Rubidium-potassium mixtures feature several intra- and inter-species collisional Feshbach resonances~\cite{inouye2004observation,klempt2007k,ferlaino2006feshbach}, where an external magnetic field can be applied to continuously tune the strength of the collisional interaction, and molecules can be adiabatically formed or dissociated without the sudden energy change typical of a chemical reaction~\cite{Timmermans1999FeshbachRI}. Despite their extremely weak binding energies on the order of nano-eV, such ``halo'' molecules can be essential resources for future AI experiments and, when more than two atoms are involved, provide fundamental new insights into quantum mechanical few-body physics~\cite{NAP25613}.
The mixtures of rubidium and potassium reported here will be used by the first teams of CAL investigators to study microgravity-enhanced molecular formation and to control differential center-of-mass distributions of $^{41}$K and $^{87}$Rb gases~\cite{Williams2017TB,PhysRevA.95.012701,Corgier2020}, while $^{39}$K will be used for high-precision investigations of Efimov physics~\cite{Cornell2017TB}. 

\begin{figure}[bp]
\centering
\includegraphics[width=0.95\linewidth]{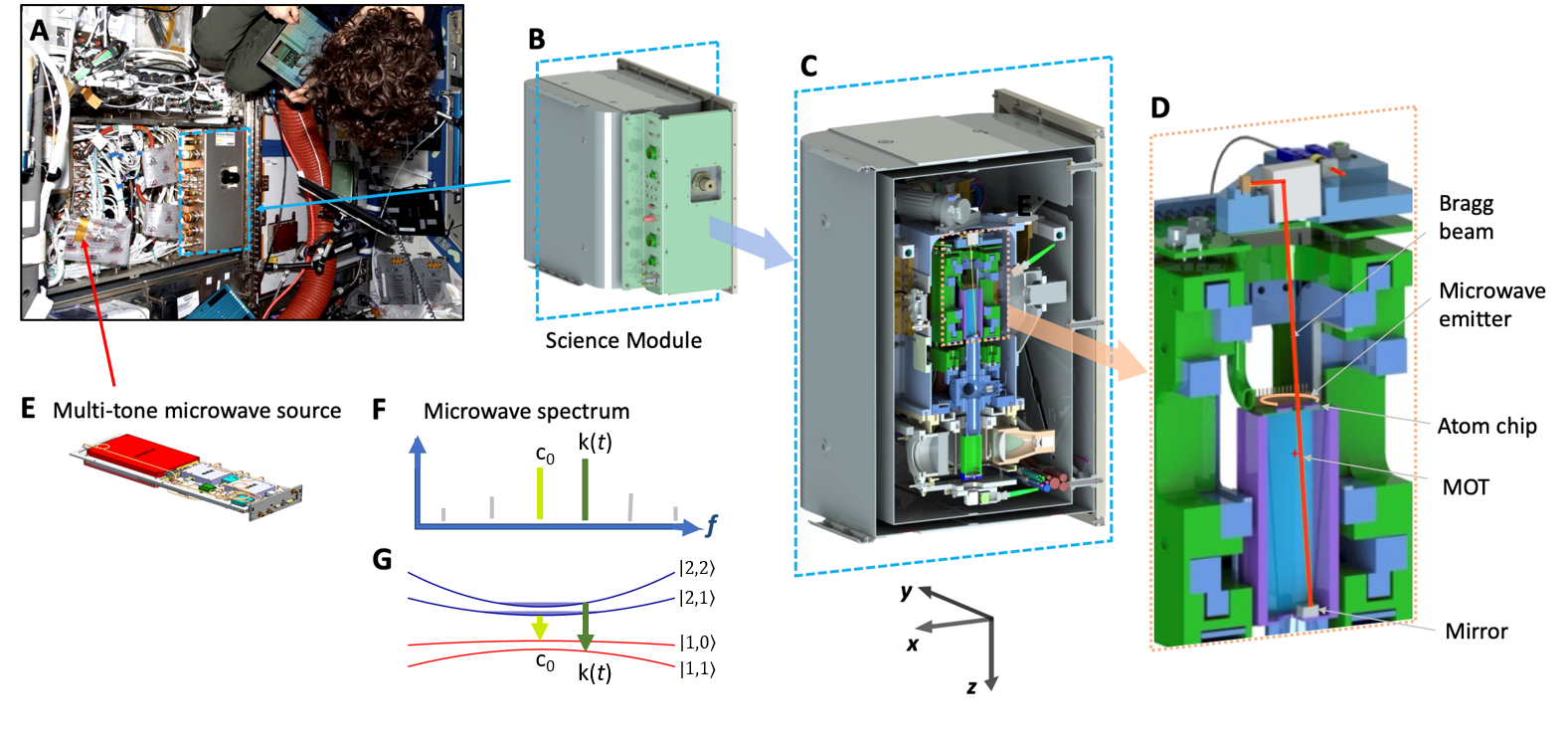}
\caption{\textbf{CAL On-orbit Hardware Upgrades:} (A) The CAL payload, housed in EXPRESS Rack 7 within the US Lab Destiny Module of the ISS, undergoing preparation for removal of the science module launched with CAL in 2018 (image source: NASA). This original science module, designated ``SM2'', was utilized for the results reported in ~\cite{CAL_nature_20,Carollo2022,CassDecomp,Gaaloul2022}), and was replaced with science module ``SM3'' by ISS crew in January 2020 (astronaut Christina Koch pictured). The novel capabilities of SM3 and the multi-tone microwave were necessary for the results reported here, and their locations are emphasized. (B) From the outside, SM2 and SM3 are virtually identical, but SM3 features a new atom chip layout and additional optics necessary for atom interferometry. (C) A cross section of SM3 revealing the interior of the vacuum chamber (``science cell'') where atom cooling and interferometry occurs (purple). (D) Enlarged view of the ``AI platform" above the science cell. The red line indicates the path of the 1-mm diameter Bragg beam operating at 785~nm, propagating approximately along the direction of Earth's gravity vector, labeled as the z-direction. The Bragg beam exits an optical fiber, passes through an optical filter, reflects off a polarizing beam splitter down through a 2~mm $\times$ 3~mm aperture in the atom chip surface which forms the upper cell wall, and finally retro-reflects from a mirror attached to the lower surface of the cell. The location of the overlapped rubidium and potassium MOT, from where atoms are magnetically transferred to the chip trap, is noted as a red cross. (E) The multi-tone microwave source installed in July 2021, featuring two synthesizers shown in red. The longest dimension is 38~cm. (F) The microwave spectrum of the multi-tone microwave source, where tones (c) and (k) are the mixed outputs of two independently tunable frequency synthesizers near 6.834~GHz and 15~MHz, yielding frequencies of order 6.834~GHz and 6.834~GHz + 15~MHz, respectively. Additional mixing harmonics are shown in gray. Tone (c) can act alone to transfer $^{87}$Rb atoms in the $\ket{2,2}$ state to state $\ket{1,1}$ to drive radiative evaporation, or it can be set to remove $^{87}$Rb atoms in the $\ket{2,1}$ state while sideband (k(t)) becomes a time dependent frequency to evaporates atoms from the $\ket{2,2}$ to $\ket{1,1}$ state. This later case is shown in (G).}
\label{fig:HW}
\end{figure}

The subsystems and layout of the CAL instrument have been described previously~\cite{CAL_nature_20,CALug}. Upgrades and repairs are commonplace in an atomic physics experiment, but CAL's unique presence in space requires the intervention of the ISS crew for any hardware reconfiguration or repair. For the present work, astronauts under guidance from a ground support team at JPL have performed two key upgrades: First, a new science module with additional capability for atom interferometry, and second, a new multi-tone microwave source for sympathetic cooling of the bosonic potassium isotopes $^{39}$K and $^{41}$K.
Figure~\ref{fig:HW} gives a visual representation of both the upgraded science module (designated as ``SM3") and the new multi-tone microwave source. We first discuss the use of the microwave source to produce ultracold potassium.

CAL has the laser facilities to generate magneto-optical traps for both Rb and K~\cite{CALug}. Because potassium is a more challenging species to laser cool, we leverage the favorable properties of $^{87}$Rb and prioritize the number of rubidium atoms loaded into our magnetic chip trap, with 
orders of magnitude fewer potassium atoms simultaneously confined as described in~\cite{CALug}. We then perform evaporative cooling on the rubidium atoms, in which the most energetic atoms are removed from the trap and the remainder allowed to rethermalize at a lower temperature. Our previous implementation of evaporative cooling worked by driving RF Zeeman transitions from trapped to untrapped spin states~\cite{CAL_nature_20,Carollo2022,Sackett2022}.  However, a feature of microwave hyperfine transitions in alkali atoms is the fact that different elements, and even different isotopes, have transitions that are well separated from each other in their transition frequency. Therefore, using the newly installed microwave source, we can selectively drive transitions in $^{87}$Rb to remove hotter rubidium atoms during the evaporative cooling process, without removing the less abundant co-trapped potassium atoms. The potassium atoms then cool down sympathetically through collisions with the actively cooled rubidium~\cite{Firstsym, SilberLi,ChineseMuWaveRbK,marzokrecentsilber,GermanMuWave,twofermiBEC,largeK39BEC,ChineseMuWaveRb,2SpeciesBEC, Waker_Tubable_K39BEC} (see Methods). Collisions between rubidium and potassium, and the efficacy of sympathetic cooling, are parameterized by an interspecies scattering length. While the scattering length between $^{87}$Rb and $^{41}$K is favorably large ($163a_0$, where $a_0$ is the Bohr radius), the scattering length between $^{87}$Rb and $^{39}$K is relatively small ($28 a_0$)~\cite{2006_K_RB_Feshbach}. Therefore, long evaporation times of tens of seconds have been required in previous Earth-based implementations to sympathetically cool $^{39}$K with $^{87}$Rb~\cite{PhysRevLett.99.010403,largeK39BEC}. Our system mitigates this problem by employing an atom chip trap that is capable of generating strong confinement, and thus a high collision rate. During the evaporative cooling process, the trap frequencies are $(\omega_{x},\omega_{y},\omega_{z})/ 2\pi =$ (26, 950, 950) Hz for $^{87}$Rb and (39, 1420, 1420) Hz for $^{39}$K, with a trap bottom at a field of 10.03~G. Although this field is far from any known interspecies Feshbach resonance, this tight confinement allows us to perform microwave evaporation in 1.68~s. As a result, our setup allows us to sympathetically cool $^{39}$K as well as $^{41}$K atoms with $^{87}$Rb using just the magnetic trap, without the need to reload into an optical trap and exploit Feshbach resonances. This simplifies the procedure and hardware --- a significant advantage for spaceborne applications. 

While the microwave cooling of $^{87}$Rb is a powerful technique, it brings additional challenges even without the presence of a second species. This is due to the fact that $^{87}$Rb atoms in both the $\ket{F, m_{F}}=\ket{2,2}$ and the $\ket{2,1}$ Zeeman states can be magnetically trapped. The uncooled $\ket{2,1}$ atoms are a source of heating, and they can collide inelastically with K which ejects both atoms from the trap~\cite{PhysRevLett.89.053202,PhysRevLett.89.190404}. A single microwave tone ramping in frequency does not automatically remove the $\ket{2,1}$ atoms, so it is necessary to separately drive both $\ket{2,2}$ and $\ket{2,1}$ states to an untrapped state using multiple frequencies. The multi-tone capability of the new microwave source accommodates this. However, spreading the microwave spectrum over multiple frequencies reduces the power available to drive any single transition, and the short wavelength can make the power coupled to the atoms both frequency and position dependent. Effective cooling therefore required careful optimization. The Methods section details the procedure we developed, which we used to produce Rb condensates with the atom number exceeding $1.5 \times 10^4$.

We first demonstrated sympathetic cooling of $^{39}$K, with natural abundance of approximately 93.3\% compared to an abundance of 6.7\% for $^{41}$K. A larger abundance generally favors the initial loading of atoms into a magneto-optical trap, thus providing a more favorable starting point for subsequent cooling steps. After using two laser systems to simultaneously capture both $^{87}$Rb and $^{39}$K in a dual species magneto-optical trap~\cite{CAL_nature_20}, we then loaded both species onto the magnetic chip.
Evaporating with the same microwave protocol used to produce a pure $^{87}$Rb BEC led to a faint $^{39}$K signal of only a few hundred atoms. However, by reoptimizing the microwave evaporation procedure (see Methods), we were able to produce $3 \times 10^4$ $^{39}$K atoms at 350~nK, as shown in Figure~\ref{fig:dual_BEC} C.
We note that $^{39}$K in the $\ket{2,2}$ state will not form a stable BEC near zero field due to the relatively strong and attractive homonuclear interactions. At present, CAL is designed to only study non-condensed ultracold samples of this species.

We then investigated cooling $^{41}$K to degeneracy.
Using the same microwave evaporation protocol for $^{39}$K, we produced the $^{41}$K BEC shown in Figure~\ref{fig:dual_BEC} B, with no $^{87}$Rb remaining. In optimizing for a binary BEC of $^{87}$Rb and $^{41}$K, we encountered the expected strong anti-correlation in atom number for each species~\cite{PhysRevA.98.063616}. With further adjustments to the evaporation sequence (see Methods), we were able to restore sufficient $^{87}$Rb population to produce the binary BEC shown in Figure~\ref{fig:dual_BEC} A. 
The final evaporation stage yields typically $1.3 \times 10^4$ degenerate rubidium atoms and $1.5 \times 10^3$ degenerate potassium atoms. With this performance, we are able to investigate the behavior of a binary quantum mixture and perform dual-species atom interferometry. 

The mean-field interaction between $^{87}$Rb and $^{41}$K atoms is repulsive at low fields, making condensates of the two species generally immiscible~\cite{PhysRevLett.89.190404}. As seen in Figure~\ref{fig:Simulation}, we observe that the condensed $^{41}$K atoms are displaced from the center of the non-condensed $^{41}$K thermal distribution when both species are present. Because the non-condensed atoms have no significant mean-field interactions compared to the condensates, this displacement is a signature of interaction effects. A similar shift has been observed in terrestrial studies of $^{41}$K and $^{87}$Rb mixtures~\cite{pichery2023efficient}. On Earth, the greater $^{87}$Rb mass results in a $^{87}$Rb condensate is typically located below the $^{41}$K condensate. Using a system of coupled Gross-Pitaevskii equations, we investigated the ground state of the two BEC components in microgravity and found it to be nearly degenerate between symmetric and asymmetric configurations of the atoms. Propagating this model further indicates that that these configurations become indistinguishable after free expansion and imaging. Rather than interaction effects impacting the initial states, we hypothesize that the observed shift occurs from interspecies interactions during the complicated dynamics of decompression and release from the atom chip trap; the z-direction being the direction along which the greatest center of mass motion occurs. The clear offset seen between the thermal and condensed components of the $^{41}$K gases in Figure~\ref{fig:Simulation}, particularly in the z-direction, provides the first evidence for interspecies interactions of degenerate matter in space. These effects will be the subject of further studies, where we aim to experimentally distinguish the ground state configuration of the trapped atoms.

To explore the feasibility of near-term quantum tests of the equivalence principle in space, we performed simultaneous atom interferometry on $^{41}$K and $^{87}$Rb BECs. CAL uses Bragg scattering~\cite{PhysRevLett.82.871} to provide the required atomic beam splitting and reflection operations, where the atomic wavefunctions interact with an optical lattice produced by off-resonant counter-propagating laser beams. A two-photon transition places the wavefunction in a superposition of different momentum states, maintaining the same internal state. By remaining in the same internal state, differential effects such as ac Stark shifts or Zeeman shifts are highly suppressed. As shown in Figure~\ref{fig:HW}, the lattice laser, or Bragg beam, enters the vacuum cell through a window at the center of the atom chip and is retro-reflected by a mirror inside the cell. The path is nominally aligned with the direction of Earth's gravity vector. The maximum laser power is 66~mW and the wavelength is 785~nm. 

Because changing the momentum of an atom changes its kinetic energy, driving a Bragg transition efficiently requires a time-dependent potential. We tailor the release of atoms from the magnetic trap in the $\ket{2,2}$ state to give the $^{87}$Rb and $^{41}$K clouds velocities of 1.2 cm/s and 1.9 cm/s, respectively, along the Bragg laser beam. In this way, the degeneracy that would otherwise allow transitions among momentum states to be driven simultaneously in two directions is broken.  
This preparation scheme results in Bragg resonance frequencies of 48~kHz for $^{87}$Rb and 88~kHz for $^{41}$K (see Extended Data). The Bragg laser is controlled by an acousto-optic modulator, which can provide three simultaneous frequency components. For single-species operation, two components are used, separated by the corresponding Bragg frequency. Using all three components, both species can be addressed at any velocity simultaneously. 

To calibrate the Bragg interaction, we measured the Rabi rate for each Bragg transition in the dual-tone configuration. We found the $\pi/2$ pulse duration to be 0.12~ms for Rb and 0.19~ms for K. 
Based on these results, we implemented single-species Mach-Zehnder interferometers using a three-pulse $\pi/2-\pi-\pi/2$ sequence with an interrogation time (defined as the time between laser pulses) of $T = 0.5$~ms. The phase of the final pulse was scanned between 0 and $2\pi$, and the final population in the two momentum states was observed to oscillate as expected, with a visibility of approximately 0.38 for $^{87}$Rb and 0.2 for $^{41}$K.

For dual species interferometry, it was necessary to recalibrate the Bragg laser in the tri-tone configuration (Extended Data). The power ratio of the frequency components was adjusted slightly to give equal Rabi rates and a shared $\pi/2$ time of 0.27~ms for both species. The three-pulse Mach-Zehnder interferometer was then applied with $T = 0.5$~ms. The results, shown in Figure~\ref{fig:Dual_AI}, exhibit clear interference in both species with measured visibilities of $0.16 \pm 0.03$ for $^{41}$K and $0.15 \pm 0.03$ for $^{87}$Rb in excellent agreement with predictions accounting for the efficiencies of each Bragg pulse (Methods). A differential phase measurement result of $0.37^{+0.45}_{-1.19}$ fulfills expectations as a starting point for future quantum tests of the UFF in space.

Building on this proof of principle demonstration, the CAL PI teams are working to realize the full performance of the instrument’s dual-species AI. Towards this goal, we note that a common Bragg beam for all atomic species within CAL gives excellent potential for suppression of common-mode noise~\cite{PhysRevA.93.013602}. The Bragg beam wavelength of 785 nm was chosen to be near “a magic wavelength” for $^{87}$Rb and $^{39}$K or $^{41}$K, where the Rabi rates of simultaneously applied Bragg pulses are equal. A shared Bragg beam then minimizes perturbations from laser frequency noise, laser intensity fluctuations, and the significant vibrations of the ISS~\cite{Williams2022}.  To the best of our knowledge, this is the first use of a "magic wavelength"for rubidium-potassium  AI, on Earth or in space. Experimental constraints that currently limit the sensor performance in terms of visibility and achievable AI interrogation time derive mainly from imperfections in the positional overlap between the Bragg laser beam and the atomic samples during AI.
The Bragg beam size, with $1/e^2$ waist $\sim 0.5$~mm, requires a) increasing precision in the release of rubidium and potassium gases with respect to the beam center, b) reduction of the expansion energy of the gases to maintain overlap of the Bragg beam profile as time of flight increases, and c) increasing environmental control over residual forces acting on the atoms.
Further, the Doppler widths of the gases are currently on-par with the maximum two-photon Rabi frequencies for Bragg transitions for both species, and there is a limited signal-to-noise in absorption imaging of clouds with thousands of atoms in extended free-fall. 
With dedicated campaigns to cool the dual-species gases to sub-nanokelvin temperatures, preparation into magnetically insensitive (e.g., $\ket{2,0}$) states, and advanced control over the cloud center of mass motion~\cite{Sackett2022,Gaaloul2022}, it is expected that fringe visibility at interrogation times in the hundreds of milliseconds will be achievable. 

In summary, we have commissioned the upgraded capabilities of the CAL instrument to form quantum gas mixtures in low-Earth orbit, including the condensation of $^{41}$K and $^{87}$Rb to form the first dual-species BECs in space, the observation of interspecies interactions, and the preparation of ultracold $^{39}$K.
We have also achieved dual-species AI in space, utilizing simultaneous Bragg interferometry of $^{87}$Rb and $^{41}$K with a single laser interrogating both species at the magic wavelength. This work is pursued by CAL investigators as a direct pathfinder for future precision measurements with atom interferometers, including quantum tests of the universality of free fall in space, spectroscopic mass measurements via the photon recoil, and the direct detection of candidate sources to explain dark-sector physics~\cite{Bigelow2015TB,Williams2017TB,Sackett2016TB}. It further provides a platform for investigating novel few- and many-body physics phenomena in the absence of gravitational perturbations.
For instance, tuning interactions via homonuclear ($^{39}$K) or heteronuclear ($^{41}$K + $^{87}$Rb) Feshbach resonances can be used to form two- or three-atom molecules with large spatial extent that are extremely weakly bound. Work is already underway on CAL to form such states in extremely low-energy regimes to test the universality of Efimov physics\cite{Cornell2017TB} and to use heteronuclear molecules as a source of atomic mixtures with essentially perfectly overlapping spatial distributions~\cite{Williams2017TB,PhysRevA.95.012701}. The accessibility to Feshbach resonances will enable CAL scientists to tune interactions to study complex behaviors of dual-species gases confined in shell-shaped traps~\cite{Lundblad2016TB}, miscible/immiscible behaviors for heteronuclear gases with tunable interactions to stabilize quantum gases in shell potentials, and the emergence of structure and complexity in the absence of gravity~\cite{Gaaloul2022,Bigelow2015TB}. In the near-term, it is anticipated that space-enabled studies of 
quantum systems will include quantum droplets~\cite{Engelsdroplets2023}, 
matter-wave interferometry with halo molecules of large spatial extent to explore the fundamental limits of superposition~\cite{ QST_Molecule_paper,PhysRevLett.74.4783, Penrose2014}, 
the collisional dynamics of interacting dual-species gases~\cite{pichery2023efficient}, 
and molecular dissociation as a source of motionally entangled atoms~\cite{QST_Molecule_paper}.
 The confluence of quantum mixtures, large molecules, matter-wave interferometry and microgravity will be a powerful method for precision metrology, quantum simulation and cosmology, and testing the foundations of modern-day fundamental physics.

\section*{Methods} 

{\bf Microwave evaporation.} 
The microwave source features a tuneable synthesizer operating in a range near 6.8~GHz, which can be mixed with the output of a second synthesizer operating at 4-50~MHz. Operating at a single tone, the maximum output power is 38~dBm.  For pure $^{87}$Rb evaporation, an initial cooling stage used a single tone to drive the $\ket{2,2}\rightarrow\ket{1,1}$ transition. In a later stage, the multi\-tone output was used with the carrier driving the $\ket{2,1} \rightarrow \ket{1,0}$ transition, while a sideband cooled the $\ket{2,2}$ state. Powers were 29~dBm in the carrier and 33~dBm in the sideband. To sympathetically cool K atoms, it was crucial to remove the $\ket{2,1}$ Rb atoms quickly and efficiently. We achieved this through a combination of periodically jumping the single-tone output to the $\ket{2,1}$ trap bottom during the first stage, and switching to the multi\-tone output sooner.

To produce dual-species condensates, it was necessary to preserve more of the Rb atoms during evaporation. We achieved this by adding an additional RF field set to be resonant with the Zeeman transition at a field about 5~G (equivalent to 300~\textmu{K}) higher in the magnetic trap than where the microwave tones were resonant. This increased the final atom number by about 20\% even for pure Rb evaporation. We attribute this benefit to the presence of hotter atoms in the trap which the microwave tone has missed, which the RF field prevents from thermalizing with the colder atoms at the trap bottom~\cite{DieckmannThesis2001, MyattThesis1997}. A second possible explanation is an effective power increase in the microwave transition due to a double-dressing of the $^{87}$Rb atoms by the microwave and RF frequencies~\cite{PhysRevA.74.053413}, but we did not directly observe any effect of the RF field on the microwave Rabi frequency. Adding the RF field also reduced the final amount of $^{41}$K, likely because it ejected some warmer K atoms from the trap.

\begin{figure}[tbp]
\centering
\includegraphics[width=0.95\linewidth]{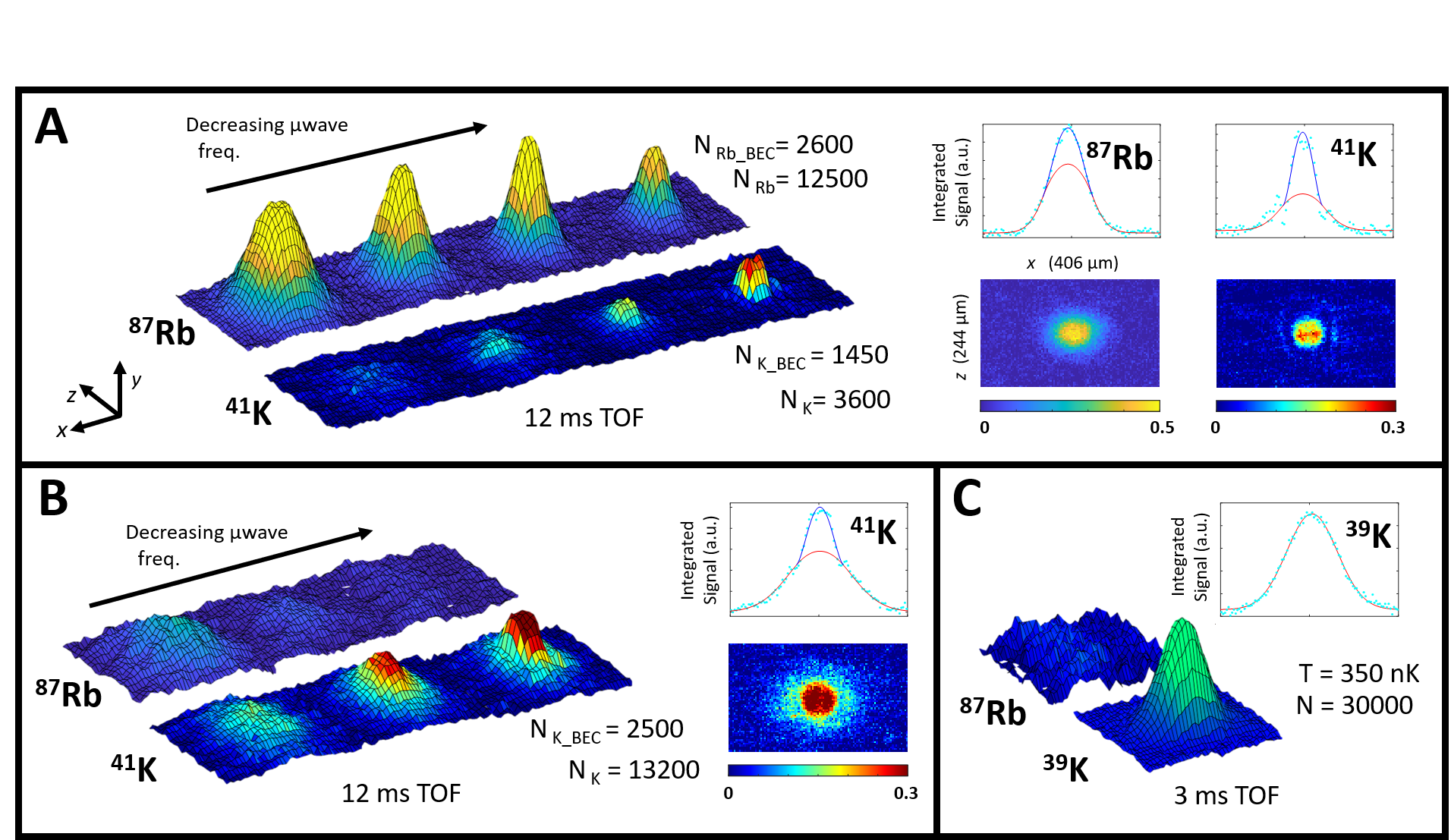} 
\caption{\textbf{Production of Degenerate Quantum Mixtures in Space:} False-colour absorption images of $^{87}$Rb, $^{41}$K, and $^{39}$K atomic clouds following microwave evaporation, decompression, and release from CAL’s magnetic chip trap (see Methods). All images are 244 by 406~{\textmu}m (60 by 100 pixels) in the $z$- by $x$-directions, respectively, where $z$ is along the direction of Earth's gravity gradient.  The potassium cloud is imaged following the indicated time of flight in free expansion (TOF), and then a separate laser frequency images $^{87}$Rb after a 2.1~ms delay. The vertical axis of all 3D surface plots is optical density (OD), and height is referenced to a common scale with a maximum of 3.5. Each atomic species then has a separate relation between height and color, where a different false color map  is assigned to each species with the color limits corresponding to the OD ranges indicated. (A) Preparation of a dual species degenerate gas. From left to right, the final frequency of an evaporative cooling ramp using a multi-tone microwave source is lowered, with the microwave evaporation supplemented by an additional RF source. In this example, $^{87}$Rb condenses first at a critical temperature at 70~nK (second from left). Further evaporation (third from left) lowers the $^{87}$Rb temperature to 59~nK, remaining below a lower critical temperature as the atom number simultaneously drops. The final evaporation step lowers the $^{87}$Rb atom number to $1.25 \times 10^4$ and temperature to 46~nK, producing a $^{41}$K BEC with $1.45 \times 10^3$ atoms at 37~nK. Temperatures are calculated from the measured BEC fraction, atom number, and trap frequencies. One dimensional density plots (light blue points), obtained by integrating along the $z$-direction (radial dimension of the trap) show the bimodal fit for the BEC component (blue) and the thermal component (red) in the final evaporation stage. (B) Potassium performance is prioritized by evaporating with only microwaves and a $^{87}$Rb BEC never forms. The atom number of $^{41}$K is maintained at $\sim 1.32 \times 10^4$ in all three images as the critical temperature of 95~nK is passed in the middle image, with a final temperature of 70~nK in the right image.  (C) $^{39}$K sympathetically cooled to ultracold temperatures, with no $^{87}$Rb remaining.}
\label{fig:dual_BEC}
\end{figure}

\begin{figure}[h]
    \centering
\includegraphics[width=0.8 \linewidth]{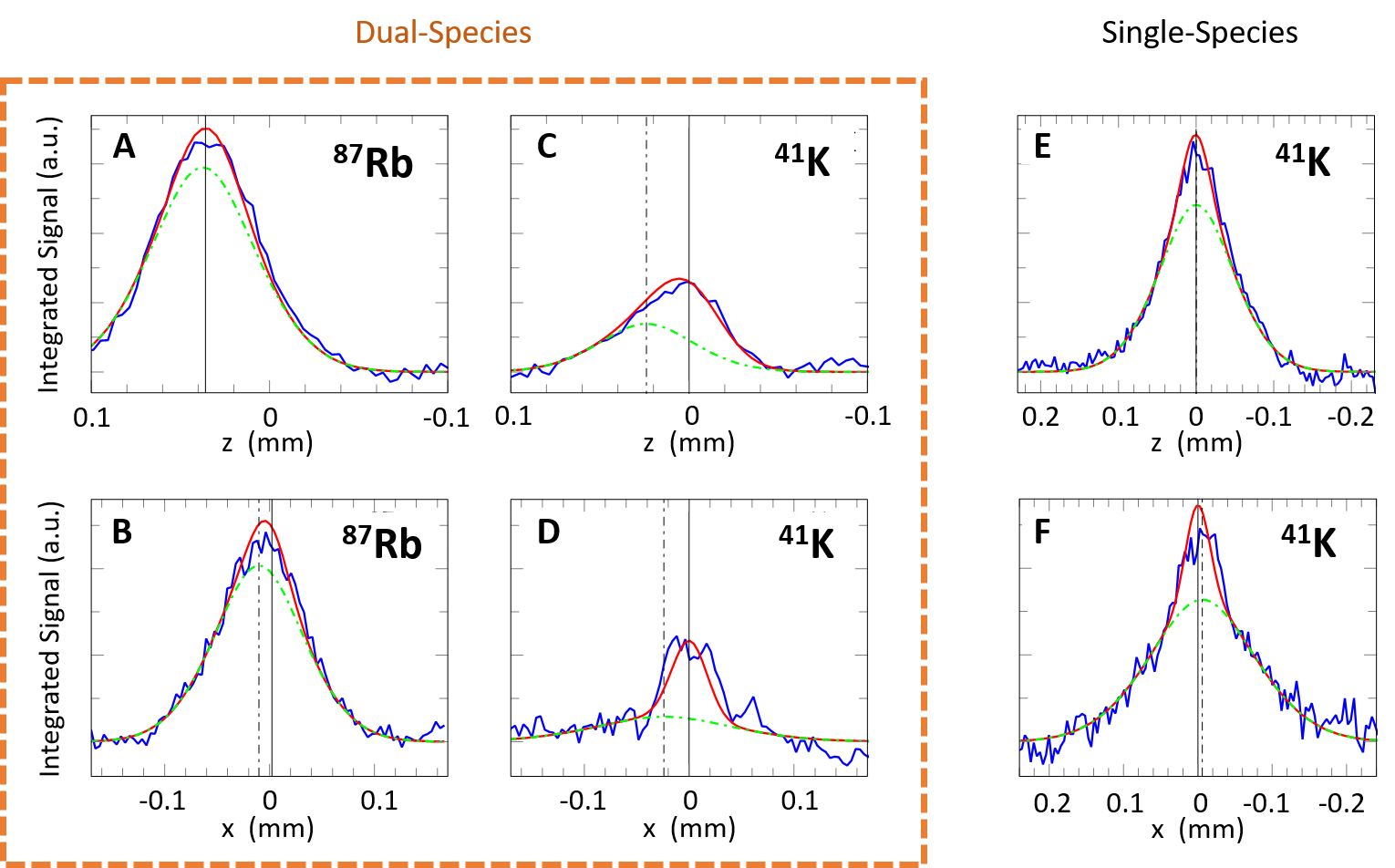}
\caption{\textbf{Interactions of degenerate $^{87}$Rb and $^{41}$K mixtures:} Blue traces are $^{87}$Rb and $^{41}$K absorption profiles from the final evaporation stages of  Figure~\ref{fig:dual_BEC} (A) and (B), integrated along the $x$- and $z$-directions respectively. Red traces are theoretical models obtained by solving for the ground state configuration of the trap, and then propagating the initial states through the processes of trap decompression, release, and free expansion. The distributions are convolved with a 15~\textmu{m}-Gaussian to account for the resolution of the camera. Position 0 is defined as the center of the condensate component for $^{41}$K, where the positive z-direction points away from the chip, consistent across figures. The solid vertical lines show the center position of the condensate component for each species. Green dashed traces are Gaussian fits to the data, showing the non-condensed atoms in the gas. The dashed vertical lines show the center of the non-condensed component. For the dual-species case, (A), (B), (C), and (D) we see that the Rb atoms share a nearly common center. In contrast, the $^{41}$K condensate is offset relative to the $^{41}$K thermal component by $\sim$30~\textmu{m} along $z$ (normal to the atom chip). This displacement is consistent with model calculations that take into account mean-field interactions between the two condensates. (E) and (F), where no $^{87}$Rb is present, displays the overlap of the $^{41}$K condensate and $^{41}$K thermal component restored.
}
    \label{fig:Simulation}
\end{figure}

\begin{figure}[tbp]
\centering
\includegraphics[width=0.49\linewidth]{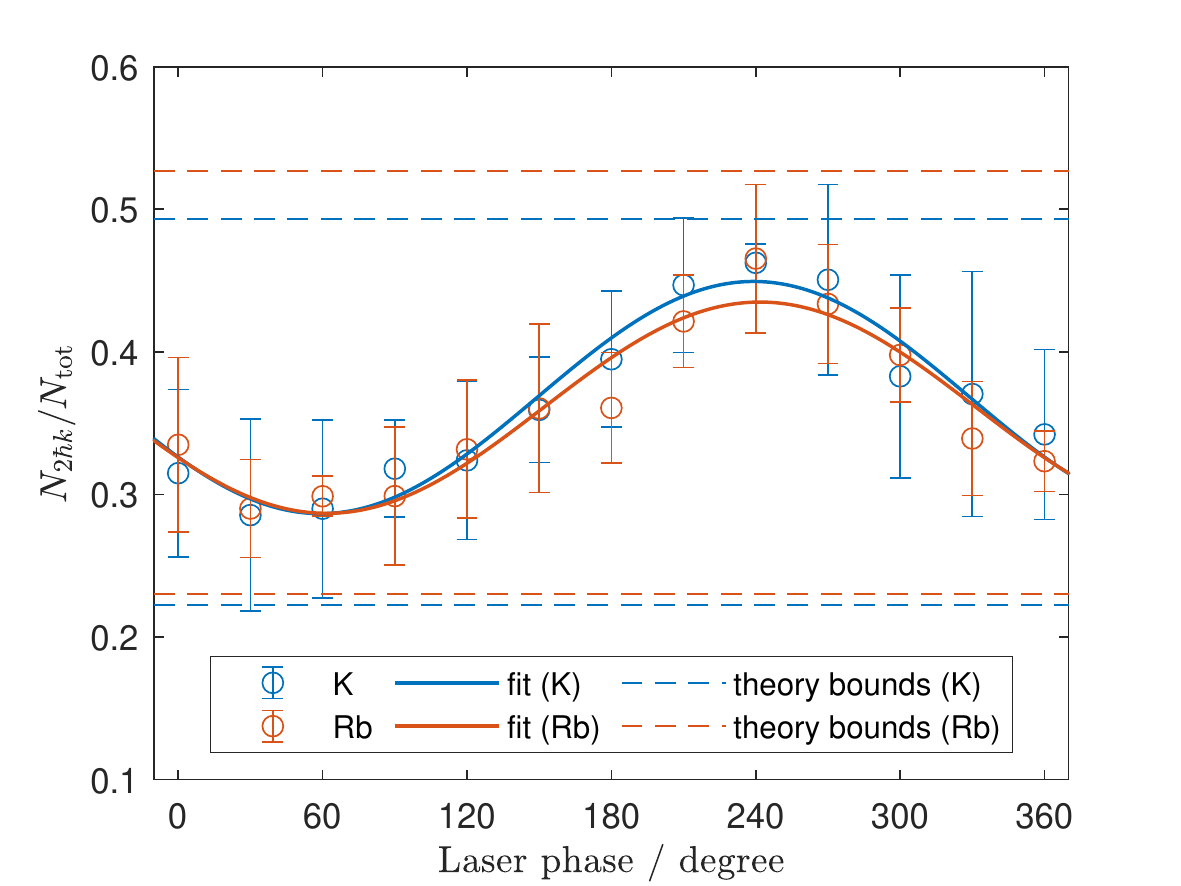}
\includegraphics[width=0.49\linewidth]{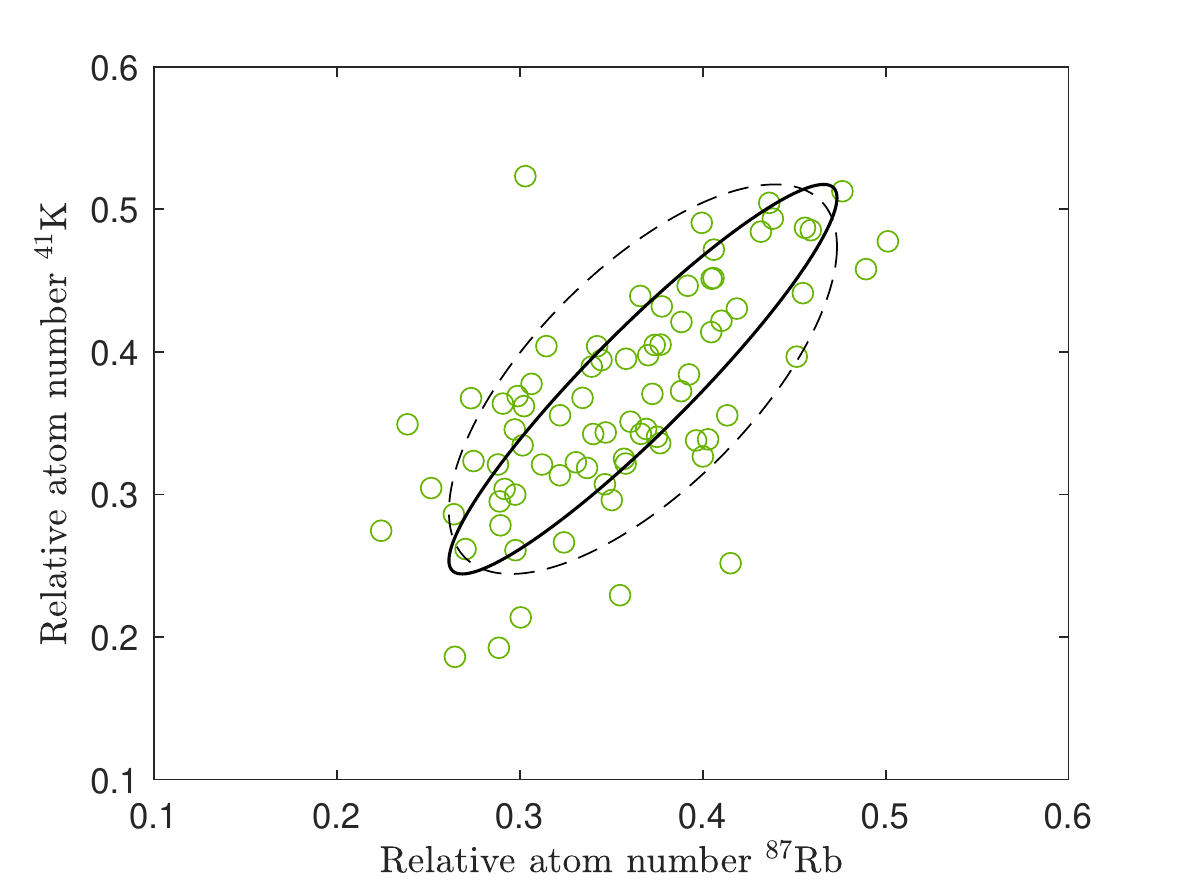}
\caption{\textbf{Dual Species Atom Interferometry in Space:} (Left) Normalized population for ultracold samples of $^{41}$K (blue) and $^{87}$Rb (red) in the excited momentum state $2\hbar k$ following the application of three Bragg pulses in a Mach-Zehnder configuration. The time between pulses is $T = 0.5$~ms, and the pulse durations are 270~\textmu{s}, 580~\textmu{s}, and 270~\textmu{s}, respectively. Each pulse interacts with both atomic species simultaneously, and contains three frequency components configured to drive Bragg transitions in both species with equal Rabi rates. For the third pulse, the same variable phase shift (degrees) is applied to components addressing each species, in steps of 30$^\circ$. Interference causes the population of the output states to vary sinusoidally, as observed. The offset and amplitude of the oscillation agrees very well with the expected minimal and maximal populations in the excited momentum state obtained by modelling the efficiencies of all three pulses (see Methods). Each data point and error bar represents the average and standard deviation of between 5 and 15 independent experimental runs respectively for a chosen phase shift.
(Right) Comparison of the relative population in the excited momentum state of $^{41}$K and $^{87}$Rb showing a correlation between both species. Fitting an ellipse (black) to the data yields a vanishing differential phase of $0.37^{+0.45}_{-1.19}$ indicating that both interferometers measured the same phase. }
\label{fig:Dual_AI}
\end{figure}

{\bf Decompression.} Evaporative cooling takes place in a tightly confining trap with $^{87}$Rb oscillation frequencies $(\omega_{x},\omega_{y},\omega_{z}) =2\pi\, \times$ (26, 950, 950)~Hz. To generate the dual species BEC images of Figure~\ref{fig:dual_BEC} and~\ref{fig:Simulation}, the evaporation trap was decompressed to a trap with $^{87}$Rb oscillation frequencies $(\omega_{x},\omega_{y},\omega_{z}) =2\pi\, \times$ (23, 83, 77)~Hz via a reduction of the $x$- and $y$-bias fields to 20\% of their evaporation value. To prepare atoms for AI, we adjust bias fields and chip currents of the evaporation trap to adiabatically displace the trap center to the position of the Bragg beam. The confinement is also reduced, leading to Rb frequencies of $(\omega_{x},\omega_{y},\omega_{z}) =2\pi\, \times$ (10, 25, 20)~Hz. Before releasing the atoms, we suddenly displace the trap so as to apply a velocity kick to the atoms. 
Following this step, we measure the center of mass velocities of the two clouds as 12.8~mm/s for Rb and 20.7~mm/s for K, and the cloud widths at 27~ms time of flight correspond to effective temperatures of 10~nK (Rb) and 4~nK (K).

{\bf Bragg beam.} The Bragg beam is controlled by a fiber coupled acousto-optic modulator (AOM), driven by the amplified output of an RF arbitrary wave form generator (AWG). This RF output drives both the AOM and the loop antenna used for RF evaporation, as selected via a switch. The AWG is a National Instruments PXI-5422, capable of a mono-, dual-, and tri-tone output in the range of 0-80~MHz. Each atomic species requires at least two tones for two-photon Bragg diffraction: One carrier tone at the AOM resonance of 79~MHz, and one offset from 79~MHz at the velocity-dependent Bragg resonance. For simultaneous AI, the tri-tone provides a 79~MHz carrier with the relevant offset Bragg resonance for each species.

{\bf Imaging system} Two imaging systems are employed in CAL, each supporting the collection of absorption and fluorescence images. The former are taken at the end of each experimental sequence for determining potassium and rubidium atom numbers and density profiles, while the latter are utilized to verify the proper operation of the MOT. The circularly polarized imaging beam can be directed either along the surface of the atom chip ($y$-axis) or through the atom chip window ($z$-axis). The 2D density profiles presented and analyzed for this paper were collected via $y$-axis absorption imaging, with resolution at the focus measured before flight to be better than 14~\textmu{m} for 780~nm light. 
The 8.25~mm field of view captures both the position of the atoms after decompression and the position within the center of Bragg beam displaced by 2~mm along the $y$-axis. However, the relatively small depth of focus of our $y$-axis imaging system (calculated to be $\sim 0.13$~mm for 780~nm light) and the inability for refocusing the cameras in flight resulted in moderate defocusing of the imaged atoms for both trap positions. 

The imaging sequence for each atomic species constitutes of two 100~\textmu{s} wide pulses that are separated by 106.3~ms. Absorption imaging of rubidium and potassium in the same experimental run interleaves these pulse sequences such that the imaging pulse for potassium is taken with 780~nm light first and the imaging pulse for potassium is taken with 767~nm light 2.1~ms later. In Figure~\ref{fig:dual_BEC}A and Figure~\ref{fig:dual_BEC}B, absorption images of the $^{41}$K and $^{87}$Rb clouds are taken after 12~ms and 14.1~ms of free expansion time respectively. $^{39}$K is imaged in Figure~\ref{fig:dual_BEC}C after 3~ms free expansion time. The optical column densities and their projections along the $x$- and $z$-axis are also depicted, and the atom numbers and cloud widths are extracted from the best two-dimensional Gaussian fit parameters. In order to increase the contrast between atom clouds and background images and account for stationary interference fringes, a reference region is selected with the cloud being left outside the field of view, to scale the background images. For the fits of Figure 3, the image quality was enhanced by a principal component analysis (PCA), which identifies and enhances the most significant structures. 

\begin{figure}[h]
\centering
\includegraphics[width=0.48\linewidth]{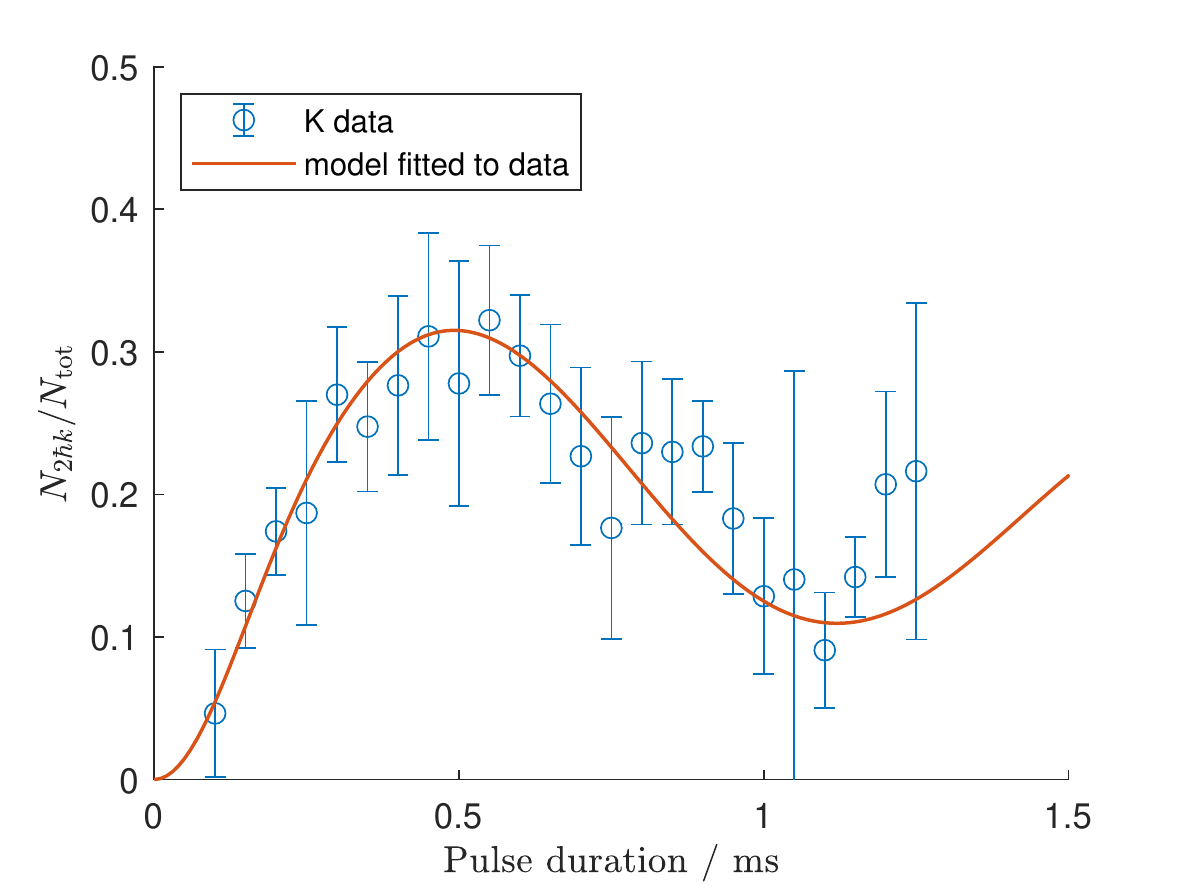}
\includegraphics[width=0.48\linewidth]{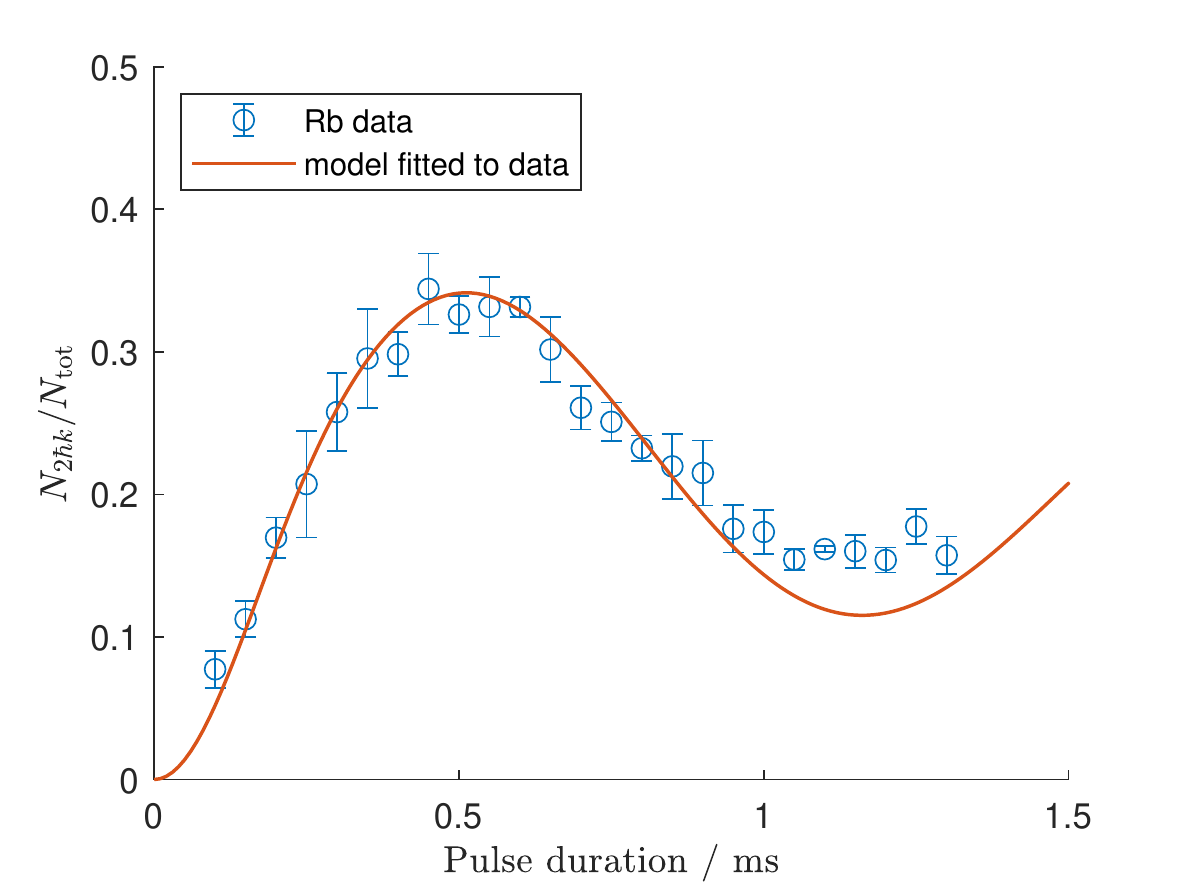}
\caption{\textbf{Rabi scan:} Relative population of the excited momentum state $2\hbar k$ of a cloud of $^{41}$K (left) and $^{87}$Rb (right) atoms as a function of the pulse duration of a single Bragg pulse. Each data point and error bar represents the average and standard deviation of 3 independent experimental runs for a chosen pulse duration. The sequence uses the tri-tone configuration that was also applied in the full interferometer shown in Figure~\ref{fig:Dual_AI}. By fitting a theory model to the data set the Rabi frequency and the momentum width of the atom cloud was determined (see Methods).}
\label{fig:Rabi_scans}
\end{figure}

\begin{figure}[h!]
\centering
\includegraphics[width=0.48\linewidth]{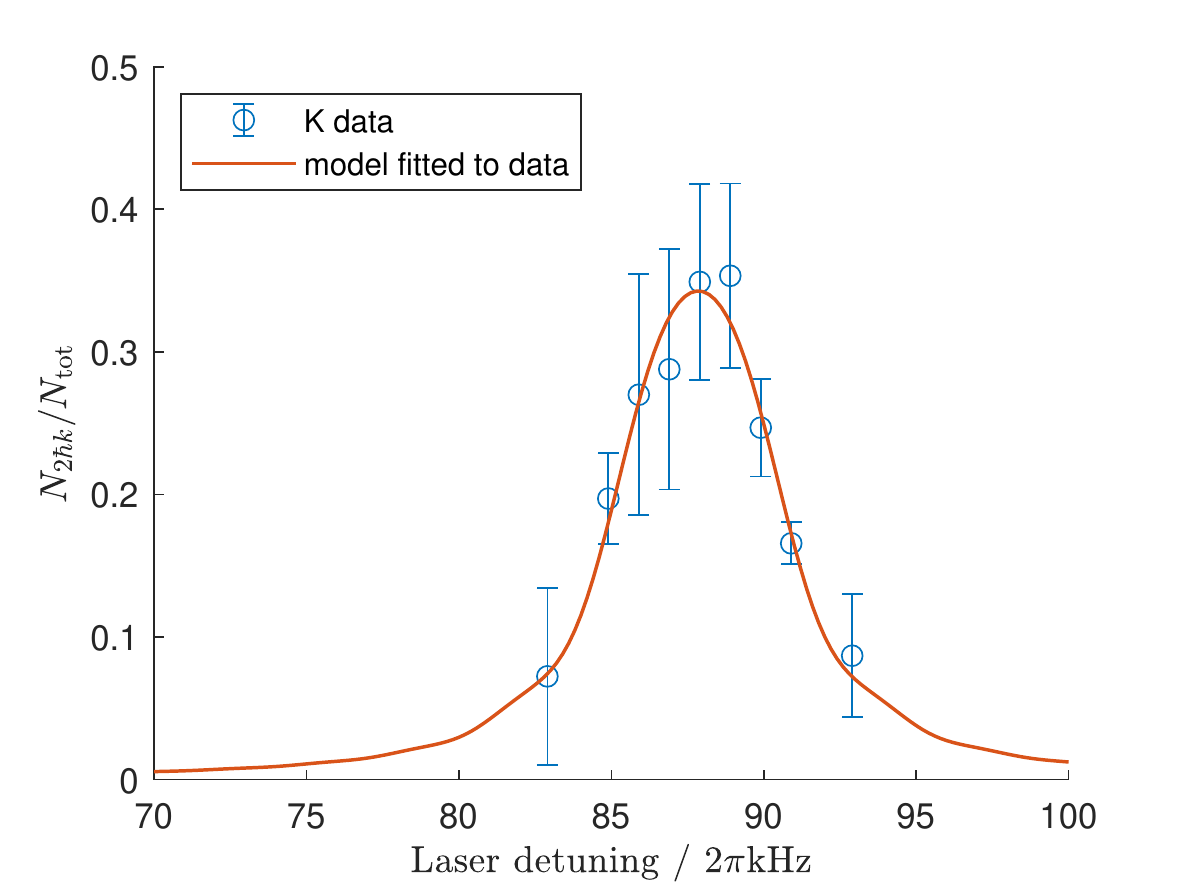}
\includegraphics[width=0.48\linewidth]{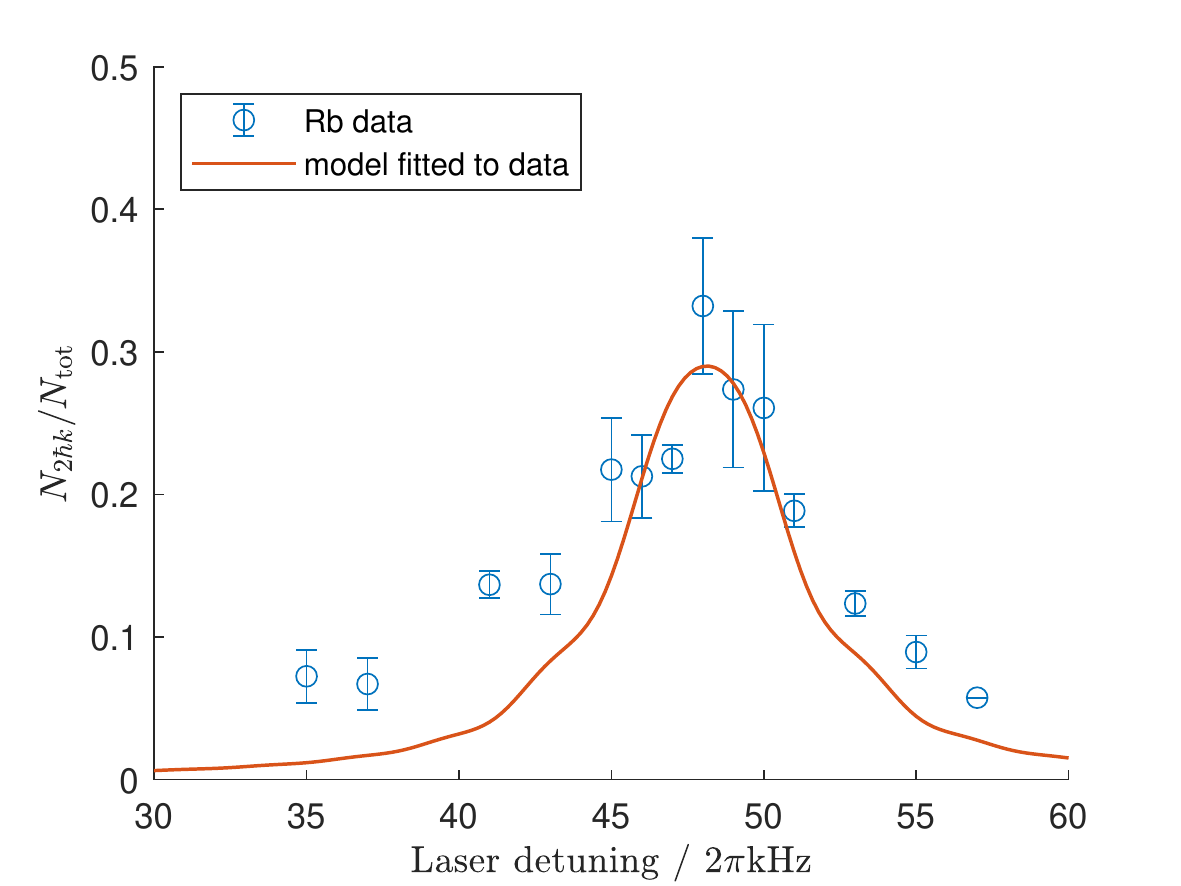}
\caption{\textbf{Detuning scan:} Relative population of the excited momentum state $2\hbar k$ of a cloud of $^{41}$K (left) and $^{87}$Rb (right) atoms as a function of the laser detuning of a single Bragg pulse. Each data point and error bar represents the average and standard deviation of 3-6 independent experimental runs for a chosen frequency. The sequence uses a two-tone configuration that delivers more laser power compared to the full interferometer shown in Figure~\ref{fig:Dual_AI}, but can only address a single atomic species in each run.  By fitting a theory model to the data set the central detuning of the atom cloud and an adjusted Rabi frequency was determined as explained in the Methods section. Due to the difference in laser power, the Rabi frequency obtained from this measurement was not used to model the full interferometer.}
\label{fig:detuning_scans}
\end{figure}

{\bf Dual species interferometry analysis.} The dual-species interferometer and the preparatory detuning and Rabi scans were analyzed by modelling the effect of the Bragg beams~\cite{Siemss_2020,Jenewein_2022} on the momentum distribution of the atom clouds and subsequently calculating the populations in the relevant exit ports as an integral over all momenta. For the model we have assumed initial Thomas-Fermi distributions in momentum space for both species and considered transitions only between the two resonantly-coupled momentum states. The Bragg pulses are approximated as square pulses. The momentum widths $\Delta p_z = 0.142 \pm 0.008 \,\hbar k$ for $^{41}$K and $\Delta p_z = 0.264 \pm 0.008 \,\hbar k$ for $^{87}$Rb as well as the Rabi frequencies $\Omega = 2\pi \cdot (786 \pm 29)$~Hz for $^{41}$K and $\Omega = 2\pi \cdot (758 \pm 15)$~Hz for $^{87}$Rb were determined by fitting the model to the Rabi scan (see Extended Data Figure~\ref{fig:Rabi_scans}), while the central momenta of the wave functions $p_0 = 1.779 \pm 0.004 \,\hbar k$ for $^{41}$K and $p_0 = 2.230 \pm 0.026 \,\hbar k$ for $^{87}$Rb were obtained through fits to the detuning scan (see Extended Data Figure~\ref{fig:detuning_scans}). Here $k = 2\pi/\lambda$ is the wave vector of the Bragg beam with wavelength $\lambda = 785$~nm.

Based on these results the expected minimum and maximum populations of the dual-species interferometer were calculated by considering the superposition of all four wave functions in each exit port and varying the phase of the final beam splitter between $0$ and $\pi$. An additional phase of $\exp(-\mathrm{i}p(\pm \Delta z)/\hbar)$ was applied to the momentum distributions of the spurious paths of the interferometer to account for their displacement $\Delta z = 2\hbar k T /m$ in position with respect to the central paths of the interferometer. Here $T = 0.5$~ms is the time between the interferometer pulses and $m$ is the mass of the atoms. The resulting bounds for the populations of the interferometer are given by $P_{\mathrm{min,K}} = 0.222^{+0.008}_{-0.006}$ and $P_{\mathrm{max,K}} = 0.493^{+0.031}_{-0.032}$ for $^{41}$K and $P_{\mathrm{min,Rb}} = 0.230^{+0.004}_{-0.009}$ and $P_{\mathrm{max,Rb}} = 0.527^{+0.015}_{-0.031}$ for $^{87}$Rb and are shown in Figure~\ref{fig:Dual_AI}. The values of the errors are based on the fit uncertainties of the Rabi frequency, central momentum, and momentum width.

\section*{Acknowledgments}

Cold Atom Lab was designed, managed, and operated by the Jet Propulsion Laboratory, California Institute of Technology, under contract with the National Aeronautics and Space Administration (Task Order 80NM0018F0581). CAL and the PI-led science teams are sponsored by the Biological and Physical Sciences Division of NASA’s Science Mission Directorate at the agency’s headquarters in Washington and by the International Space Station Program at NASA’s Johnson Space Center in Houston. N.G, E.M.R., W.P.S., P.B., G. M., and A. P. acknowledge support by the DLR Space Administration with funds provided by the Federal Ministry for Economic Affairs and Climate Action (BMWK) under grant numbers DLR 50WM2245-A/B (CAL-II). The authors would like to thank the entire NASA/JPL Cold Atom Lab team. Any opinions, findings, and conclusions or recommendations expressed in this article are those of the authors and do not necessarily reflect the views of the National Aeronautics and Space Administration.

\bibliography{GTB}
\bibliographystyle{naturemag}

\clearpage

\section*{Extended Data}

\end{document}